\documentclass{article}
\usepackage{amsfonts}
\usepackage{amsmath}
\usepackage{graphicx}
\usepackage{cite}
\usepackage{hyperref}

\begin{document}

\title{On the formation of the $1:2$ resonance in oscillator dynamics}
\author{Jan Kyzio\l , Andrzej Okni\'nski \\
Politechnika \'Swi\c{e}tokrzyska, Al. 1000-lecia PP 7, \\
25-314 Kielce, Poland}
\maketitle

\begin{abstract}
The dynamics of nonlinear oscillators are investigated. We study the formation of $1:2$ 
resonance in nonlinear periodically forced oscillators due to
period doubling of the primary $1:1$ resonance, or born independently. We
compute the amplitude-frequency implicit function, the steady-state asymptotic
solution, for the effective equation approximating coupled
oscillators. Working in the framework of differential properties of implicit
functions, we demonstrate that birth of $1:2$ resonances corresponds to
singular isolated points of the implicit functions. We provide numerical
examples illustrating our theoretical findings. 

\end{abstract}

\section{Introduction}

\label{Intro}

Recently, we have investigated metamorphoses of $1:2$ resonance and its
interaction with the primary resonance in the asymmetric Duffing oscillator %
\cite{Kyziol2024}. The study documented very complicated dynamics of the $1:2$
resonance. This work studies the $1:2$ resonance formation in
nonlinear periodically forced oscillators. The $1:2$ resonance is due to
period doubling of the main $1:1$ resonance or is born independently. The
latter phenomenon has not yet been investigated and deserves a separate
study.

We consider an effective equation describing approximately the dynamics of coupled
oscillators, and its special case, the Duffing equation.

Coupled oscillators can model dynamics encountered in mechanics, chemistry,
electronics, and neuroscience, see \cite%
{Mahmoud2004,Pikovsky2015,Schultheiss2011,Awal2019,Hajjaj2019,
Kozlowski1995,Kuznetsov2009,Wiercigroch2009,Perkins2012,Sabrathinam2013,Zulli2016,Luo2017,Karahan2017,Papangelo2019}
and references therein. A generic example is a dynamic vibration absorber,
consisting of a mass $m_{2}$, attached to the main vibrating system of mass $%
m_{1}$ \cite{DenHartog1985,Oueini1999} and governed by equations%
\begin{equation}
\left. 
\begin{array}{l}
m_{1}\ddot{x}_{1}-V_{1}\left( \dot{x}_{1}\right) -R_{1}\left( x_{1}\right)
+V_{2}\left( \dot{x}_{2}-\dot{x}_{1}\right) +R_{2}\left( x_{2}-x_{1}\right)
=f\cos \left( \omega t\right)  \\ 
m_{2}\ddot{x}_{2}-V_{2}\left( \dot{x}_{2}-\dot{x}_{1}\right) -R_{2}\left(
x_{2}-x_{1}\right) =0%
\end{array}%
\right\}   \label{general}
\end{equation}
where $V_{1}$, $R_{1}$ and $V_{2}$, $R_{2}$ are (nonlinear) force of
internal friction and (nonlinear) elastic restoring force for mass $m_{1}$
and mass $m_{2}$, respectively.

\section{Approximate effective equation and the Duffing equation}

\label{effective}

In what follows we make a simplifying assumption
\begin{equation}
R_{1}\left( x_{1}\right) =-\alpha _{1}x_{1},\ V_{1}\left( \dot{x}_{1}\right)
=-\nu _{1}\dot{x}_{1}.  \label{R1V1}
\end{equation}

Now, in new variables, $x\equiv x_{1}$, $y\equiv x_{2}-x_{1}$, we eliminate
variable $x$ to obtain the following exact equation for relative motion \cite%
{Okninski2006,Kyziol2013b}%
\begin{equation}
\left( M\tfrac{d^{2}}{dt^{2}}+\nu \tfrac{d}{dt}+\alpha \right) \left( \mu 
\ddot{y}-V_{e}\left( \dot{y}\right) -R_{e}\left( y\right) \right) +\epsilon
m_{e}\left( \nu \tfrac{d}{dt}+\alpha \right) \ddot{y}=F\cos \left( \omega
t\right) ,  \label{4th}
\end{equation}
where $m\equiv m_{1}$, $m_{e}\equiv m_{2}$, $\nu =\nu _{1}$, $\alpha =\alpha
_{1}$, $M=m+m_{e}$, $F=m_{e}\omega ^{2}f$, $\mu =mm_{e}/M$, $\epsilon
=m_{e}/M$ and $R_{e}\equiv R_{2}$, $V_{e}\equiv V_{2}$.

In the present work, we put
\begin{equation}
R_{e}\left( y\right) =\alpha _{e}y-\gamma _{e}y^{3},\quad V_{e}\left( \dot{y}%
\right) =-\nu _{e}\dot{y},  \label{ReVe}
\end{equation}%
and assume that $\epsilon m_{e}$, $\nu $, $\alpha $ are small, and,
accordingly, the term proportional to $\epsilon m_{e}$ can be neglected.

Introducing nondimensional time $\tau $ and rescaling variable $y$%
\begin{equation}
\tau =t\bar{\omega},\ \bar{\omega}=\sqrt{\frac{\alpha _{e}}{\mu }},\ z=y%
\sqrt{\frac{\gamma _{e}}{\alpha _{e}}},  \label{Ndim1}
\end{equation}%
we get the approximate effective equation \cite{Okninski2006}%
\begin{equation}
\frac{d^{2}z}{d\tau ^{2}}+h\frac{dz}{d\tau }-z+z^{3}=-\gamma \frac{\Omega
^{2}}{\sqrt{\left( \Omega ^{2}-a\right) ^{2}+H^{2}\Omega ^{2}}}\cos \left(
\Omega \tau +\delta \right) ,  \label{eff}
\end{equation}%
where $\gamma \equiv G\dfrac{\kappa }{\kappa +1}$, $\tan \delta =\frac{%
\omega \nu }{M\left( \omega ^{2}-\omega _{0}^{2}\right) }=\frac{H\Omega }{%
\Omega ^{2}-\Omega _{0}^{2}}=\frac{H\Omega }{\Omega ^{2}-a}$, and where
nondimensional quantities are given by

\begin{equation}
\left. 
\begin{array}{l}
h=\frac{\nu _{e}}{\mu \bar{\omega}}\text{,}\ H=\frac{\nu }{M\bar{\omega}}%
\text{,}\ \Omega =\frac{\omega }{\bar{\omega}}\text{,}\ \Omega _{0}=\frac{%
\omega _{0}}{\bar{\omega}}\text{,}\ \omega _{0}=\sqrt{\frac{\alpha }{M}}%
\text{, } \\ 
G=\frac{1}{\alpha _{e}}\sqrt{\frac{\gamma _{e}}{\alpha _{e}}}f\text{,}\
\kappa =\frac{m_{e}}{m}\text{,}\ a=\frac{\alpha \mu }{\alpha _{e}M}\text{.}%
\end{array}%
\right\}  \label{Ndim2}
\end{equation}

For $a=H=0$ Eq. (\ref{eff}) reduces to the Duffing equation with $\delta =0$

\begin{equation}
\frac{d^{2}z}{d\tau ^{2}}+h\frac{dz}{d\tau }-z+z^{3}=-\gamma \cos \left(
\Omega \tau \right) .  \label{Duffing}
\end{equation}

\section{Asymptotic solution of Eq. (\protect\ref{eff}) for the $1:2$
resonance}

We applied the Krylov-Bogoliubov-Mitropolsky (KBM) perturbation approach 
\cite{Nayfeh2011} to the rescaled effective equation (\ref{eff}) proceeding as in \cite{Janickia,Janickib}, 
obtaining
for the $1:2$ resonance of form%
\begin{equation}
z\left( \tau \right) =A_{0}+A\cos \left( \tfrac{1}{2}\Omega \tau +\tfrac{1}{2%
}\delta +\varphi \right)  \label{resonance}
\end{equation}%
the following solution
\begin{subequations}
\label{SOL}
\begin{eqnarray}
\tfrac{3}{2}A_{0}A^{2}+A_{0}^{3}+\tfrac{3}{2}A_{0}C^{2}-A_{0}+\tfrac{3}{4}%
CA^{2}\cos \left( 2\varphi -\delta \right) &=&0  \label{s1} \\
\tfrac{1}{2}hA\Omega -3A_{0}CA\sin \left( 2\varphi -\delta \right) &=&0
\label{s2} \\
\tfrac{1}{4}A\Omega ^{2}+A-3A_{0}^{2}A-\tfrac{3}{2}C^{2}A-\tfrac{3}{4}%
A^{3}-3A_{0}CA\cos \left( 2\varphi -\delta \right) &=&0  \label{s3}
\end{eqnarray}%
where (we assume that the denominators do not vanish) 
\begin{equation}
C=-\gamma \dfrac{\Omega ^{2}}{\sqrt{\left( \Omega ^{2}-a\right)
^{2}+H^{2}\Omega ^{2}}}\dfrac{1}{\frac{3}{4}A^{2}-\Omega ^{2}-1}  \label{s4}
\end{equation}
\end{subequations}

We eliminate the phase $2\varphi -\delta $\ and compute $A_{0}^{2}$
obtaining the following, rather complicated, implicit function $F$ of
variables $\Omega ,\ A$ and parameters $h$, $a$, $H$, $\gamma $ 
(assuming that the denominators do not vanish)
\begin{equation}
\left. 
\begin{array}{l}
F\left( \Omega ,A;h,a,H,\gamma \right)
=A_{0}^{4}+c_{2}A_{0}^{2}+c_{0}=0\medskip  \\ 
c_{2}=\tfrac{3}{4}A^{2}+\tfrac{3}{2}C^{2}-1,\ c_{0}=-\tfrac{3}{16}%
A^{4}+\left( \tfrac{1}{16}\Omega ^{2}-\tfrac{3}{8}C^{2}+\tfrac{1}{4}\right)
A^{2}\smallskip  \\ 
A_{0}^{2}=\frac{\Omega ^{4}+\left( 8-15A^{2}+4h^{2}-12C^{2}\right) \Omega
^{2}+6C^{2}\left( 6C^{2}-8+15A^{2}\right) +4\left( 3A^{2}-4\right) \left(
3A^{2}-1\right) }{12\left( 2\Omega ^{2}+3A^{2}+18C^{2}-4\right) }\smallskip 
\\ 
C=-\gamma \frac{\Omega ^{2}}{\sqrt{\left( \Omega ^{2}-a\right)
^{2}+H^{2}\Omega ^{2}}}\frac{1}{\frac{3}{4}A^{2}-\Omega ^{2}-1}%
\end{array}%
\right\}   \label{F(Omega,A)}
\end{equation}

\section{Singular points of implicit function (\protect\ref{F(Omega,A)})}

Singular points of the implicit function $F\left( \Omega ,A;h,a,H,\gamma
\right) =0$ are given by \cite{Fikhtengolts2014,Kyziol2024a} 
\begin{subequations}
\label{SING1}
\begin{eqnarray}
F\left( \Omega ,A;h,a,H,\gamma \right) &=&0  \label{sing1} \\
\frac{\partial F\left( \Omega ,A;h,a,H,\gamma \right) }{\partial A} &=&0
\label{sing2} \\
\frac{\partial F\left( \Omega ,A;h,a,H,\gamma \right) }{\partial \Omega }
&=&0  \label{sing3}
\end{eqnarray}
\end{subequations}
If we assume, for example, values of $h$, $a$, and $H$, we can solve Eqs. (\ref%
{SING1}) for $\Omega $, $A$, $\gamma $ numerically.

We can also consider a special case of singular points with $A=0$. The
corresponding conditions read
\begin{subequations}
\label{SING2}
\begin{eqnarray}
F\left( \Omega ,0;h,a,H,\gamma \right) &=&0  \label{sing_a} \\
\frac{\partial F\left( \Omega ,0;h,a,H,\gamma \right) }{\partial \Omega }
&=&0  \label{sing_b}
\end{eqnarray}%
\end{subequations}
since $\left. \dfrac{\partial F\left( \Omega ,A;h,a,H,\gamma \right) }{%
\partial A}\right\vert _{A=0}\equiv 0$.

Equations (\ref{SING2}) can be solved for $h,\ \gamma $ yielding two
polynomial equations with coefficients depending on $\Omega $, $a$, $H$, see Appendix \ref{SE}. 
Alternatively, we assume values of $\Omega $, $a$, $H$, solve
Eqs. (\ref{SING2}) numerically and choose physical solutions $(h>0$, $\gamma 
$ -- real$)$.

In the case of the Duffing equation, $a=H=0$, equations (\ref{SING2}) (or (\ref{p}), (\ref{q}))
can be simplified significantly
\begin{subequations}
\label{DD}
\begin{gather}
f\left( \Omega ,h\right) =27\Omega ^{8}+\left( 252h^{2}-486\right) \Omega
^{6}+\left( 2259-2400h^{2}+500h^{4}\right) \Omega ^{4}  \label{D1} \\
+\left( 1224+2484h^{2}+200h^{4}\right) \Omega
^{2}+20h^{4}-624h^{2}-16\,128=0\smallskip   \nonumber \\
\hspace{-50pt}g\left( \Omega ,h,\gamma \right) =\left( 69h^{2}-126\right)
\Omega ^{6}+\left( 230h^{4}-1368h^{2}+1197\right) \Omega ^{4}  \label{D2} \\
+\left( 76h^{4}+789h^{2}-387\right) \Omega ^{2}+6h^{4}-654h^{2}-9000+\left(
828h^{2}+108\right) \gamma ^{2}=0  \nonumber
\end{gather}
\end{subequations}

We are mainly interested in singular points which are isolated points of the
implicit function $F\left( \Omega,A;h,a,H,\gamma \right) =0$. This is
because isolated points with $A\neq 0$ are solutions of Eqs. (\ref%
{SING1}), correspond to the birth of the $1:2$ resonance (in all
investigated cases out of chaos), while singular points with $A=0$,
solutions of Eqs. (\ref{SING2}), correspond to the birth of the $1:2$
resonance due to period doubling of the main $1:1$ resonance.

\section{Examples of birth of $1:2$ resonances}

\subsection{The Duffing equation}

To study the Duffing equation (\ref{Duffing})\ we put $a=H=0$ in equations (%
\ref{SING1}), (\ref{SING2}). Moreover, we assume, arbitrarily, $h=0.7$.

Equation (\ref{D1}), $f\left( \Omega ,0.7\right) =0$, has only two real
roots, $\Omega =\pm 2.\,617\,420$. Then the equation (\ref{D2}), $g\left(
2.\,617\,420,0.7,\gamma \right) =0$, yields $\gamma =\pm 4.\,737\,197$.
Therefore, for $h=0.7$, $\gamma =4.\,737\,197$ the singular point of the
Duffing implicit function $F\left( \Omega ,A;h,0,0,\gamma \right) =0$ arises
-- this is an isolated point $\left( \Omega ,A\right) =\left(
2.\,617\,420,0\right) $.

We now solve numerically Eqs. (\ref{SING1}) for $h=0.7$, $a=0$, $H=0$. We
obtain, of course, the previous solution, and  $\left(
\Omega ,A_{\pm }\right) =\left( 1.\,358\,480,\pm 0.813\,037\right) $ for $%
\gamma =2.\,168\,300$. 

\begin{figure}[h!]
\center
\includegraphics[width=11cm, height=7cm]{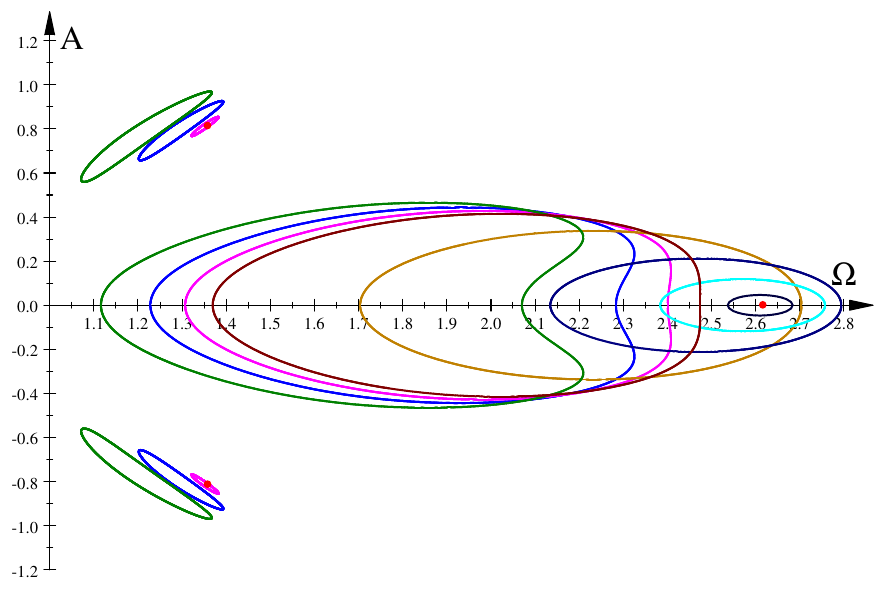}
\caption{Sequential metamorphoses of amplitude-frequency implicit function 
$F\left( \Omega ,A;h,0,0,\gamma \right) =  0$, describing $1:2$ resonance; 
$\gamma =2.15$ (Magenta), $\gamma =2.274$ (Red), $\gamma =3$ (Sienna), $%
\gamma =4$ (Blue), $\gamma =4.5$ (LtBlueGreen), $\gamma =4.7$ (Navy).
}
\label{F1}
\end{figure}

All computed singular points are isolated points and are shown in the plot below; 
see red dots in figure \ref{F1}.

For decreasing $\gamma$, the first singular point appears at $\gamma =4.\,737\,197$. 
In this isolated point $A=0$ and, therefore, corresponds to first period doubling of the 
main $1:1$ resonance. 

Then, at $\gamma =2.\,168\,300$ a pair of singular isolated points is
created, $\Omega =1.\,358\,480$, $A_{\pm }=\pm 0.813\,037$. Since $A\neq 0$,
these isolated points correspond to birth of two branches of $1:2$ resonance,  without 
a contact with the main resonance.

\newpage

To demonstrate the role of the singular points, we have computed bifurcation
diagrams for $h=0.7$ and variable $\gamma $.

\begin{figure}[h!]
\center
\includegraphics[width=6cm, height=6cm]{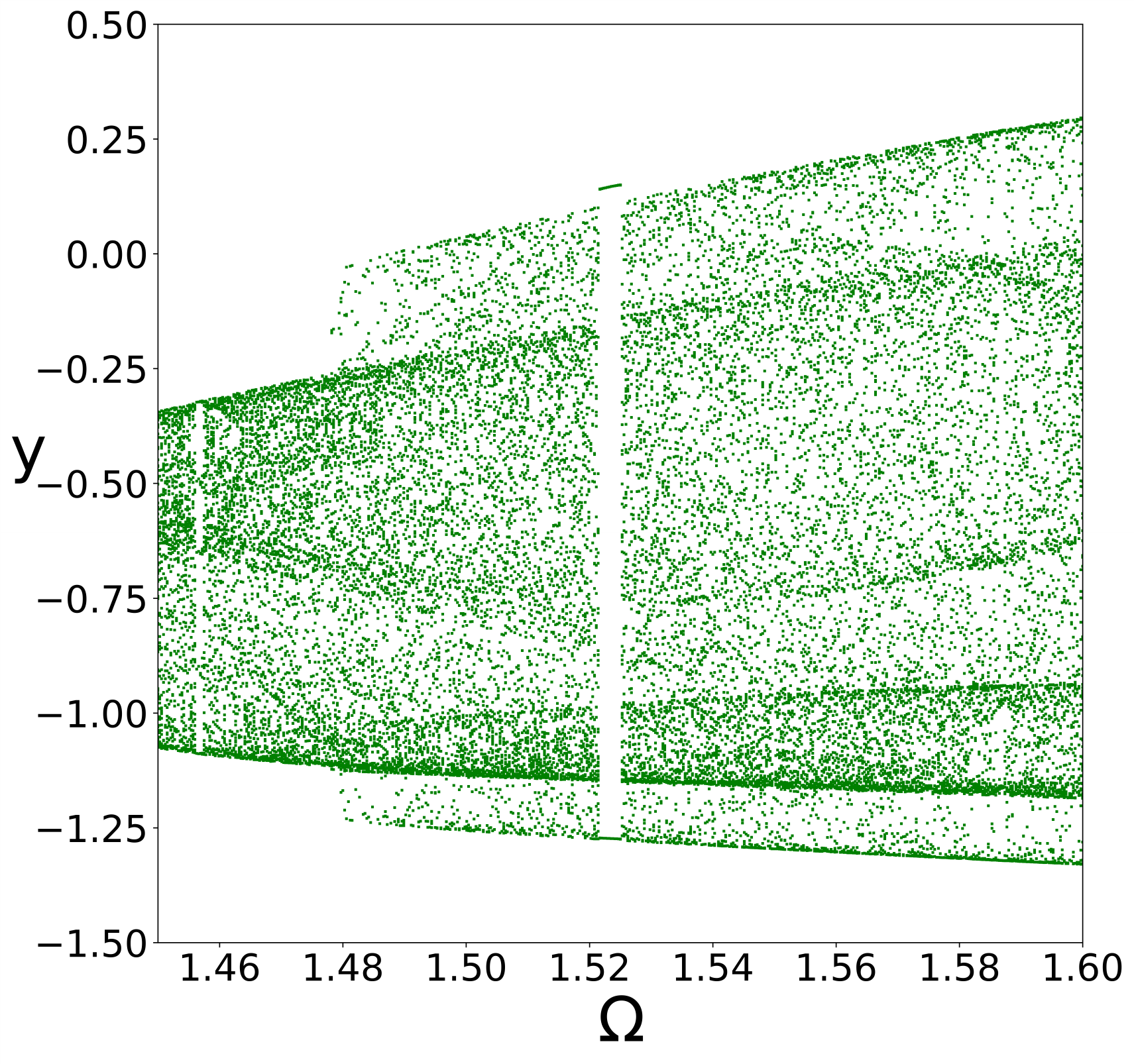}
\includegraphics[width=6cm, height=6cm]{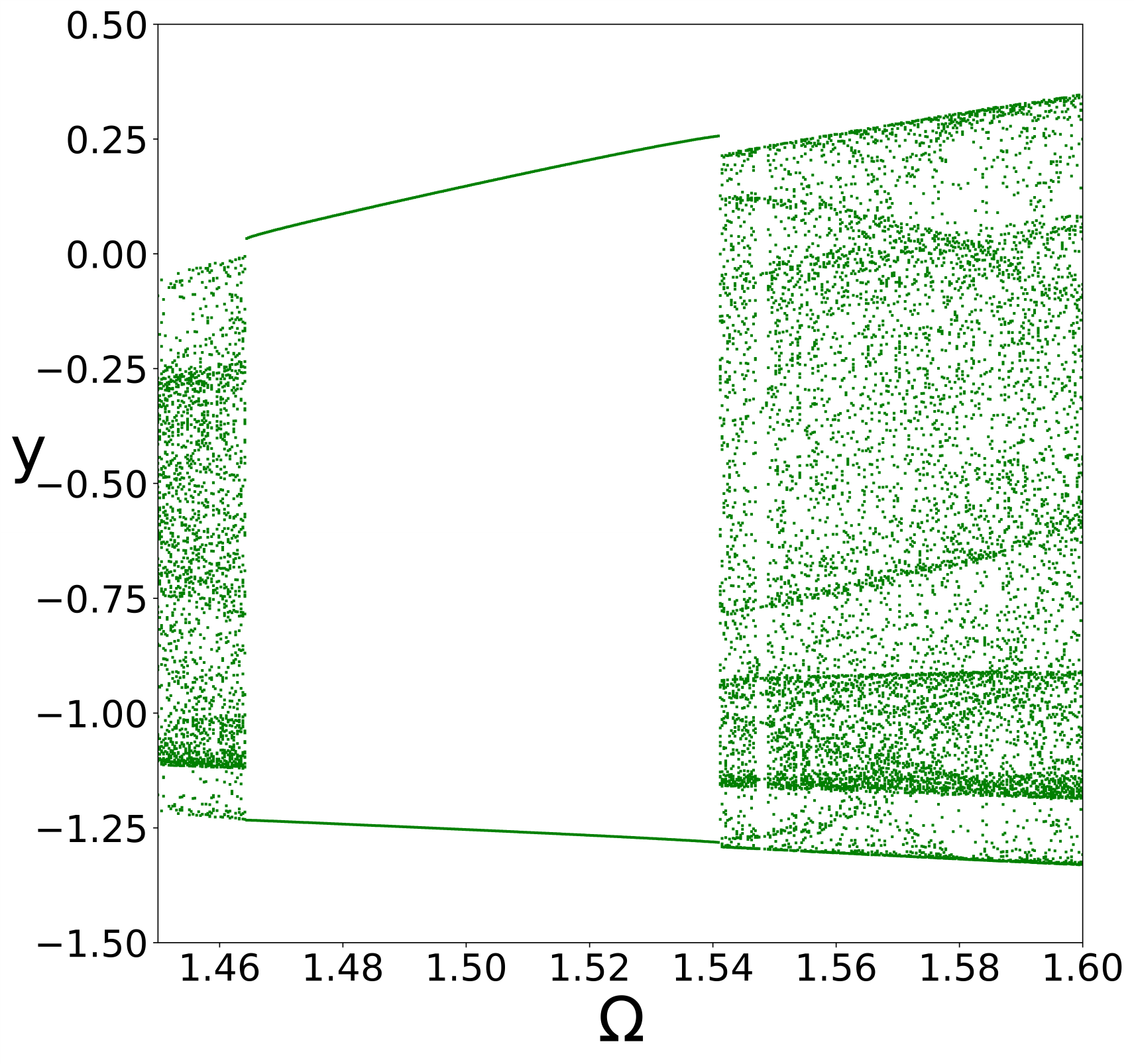}
\caption{Bifurcation diagrams: $\gamma = 1.9978$ -- left figure,  $\gamma = 1.95$ -- right figure.
}
\label{F2}
\end{figure}

Left-hand figure \ref{F2} shows birth of $1:2$  resonance out from chaos. Right-hand figure  \ref{F2} 
displays a fully developed  $1:2$ resonance. The resonance appears at  $\gamma = 1.9978$, in 
qualitative agreement with the computed value  $\gamma =2.\,168\,300$. 

\begin{figure}[h!]
\center
\includegraphics[width=11cm, height=7cm]{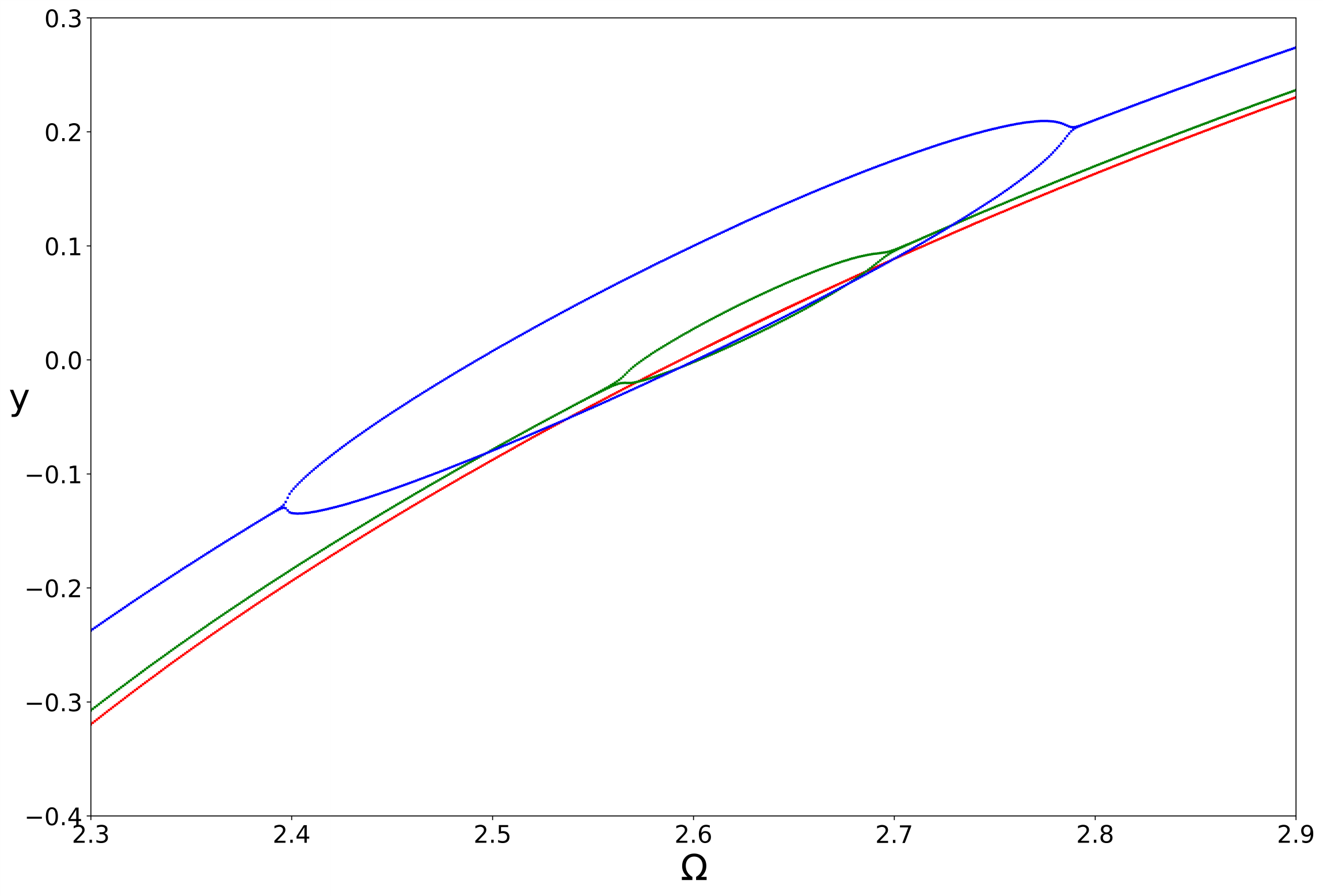}
\caption{Bifurcation diagram, $\gamma = 4.21$ -- Red,  $\gamma = 4.18$ -- Green,  $\gamma = 4.00$ -- Blue.
}
\label{F3}
\end{figure}

Fig. \ref{F3} describes period doubling of $1:1$ resonance. Red curve corresponds to 
the $1:1$ resonance just before the first period doubling at $\gamma > 4.21$. 
Green and blue curves show growth of the $1:2$ resonance (before the next period doubling). 

\subsection{The effective equation}

We consider an effective equation with, for example, $a=-0.8$, $H=0.5$, and
assume, as before, $h=0.7$.

We now solve Eqs. (\ref{SING1}) numerically, obtaining several solutions.
Namely, we get $\gamma =3.\,159\,196$ and $\left( \Omega ,A\right) =\left(
1.\,001\,811,\pm 0.527\,685\right) $ -- self-intersections, as well as $%
\gamma =3.\,\allowbreak 243\,191$ and $\left( \Omega ,A\right) =\left(
1.\,295\,330,\pm 0.742\,884\right) $ -- a pair of isolated points.

We solve Eqs. (\ref{SING2}) numerically, obtaining again several solutions.
There is a solution $\gamma =5.\,375\,442$, $\left( \Omega ,A\right) =\left(
2.\,581\,157,0\right) $, corresponding to an isolated point. Figure below
shows all singular points (red dots). 

The first singular point appears, for decreasing $\gamma $, at $\gamma
=5.\,375\,442$. In this isolated point $A=0$ and, therefore, corresponds to
the first period doubling of the main $1:1$ resonance. 

Then, at $\gamma =3.\,\allowbreak 243\,191$ a pair of singular isolated
points is created, $\Omega =1.\,295\,330$, $A_{\pm }=\pm 0.742\,884$. Since $%
A\neq 0$, these isolated points correspond to birth of two branches of $1:2$
resonance, without a contact with the main resonance. 

There is also a pair of self-intersections for $\gamma =3.\,159\,196$, unrelated, 
however, to the birth of $1:2$ resonance.

\begin{figure}[ht!]
\center
\includegraphics[width=11cm, height=7cm]{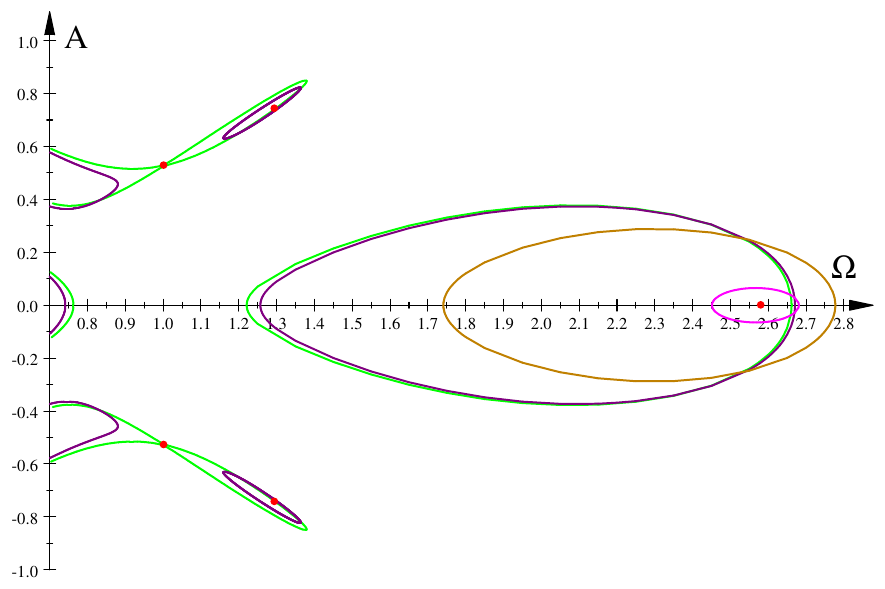}
\caption{Amplitude-frequency implicit function  $F\left( \Omega ,A;h,a,H,\gamma
\right) =0$: 
$\gamma =5.3$ (Magenta), $\gamma =4$ (Sienna), $\gamma =3.20$ (Purple), $%
\gamma =3.159$ (LightGreen).
}
\label{F4}
\end{figure}

We have computed bifurcation diagrams for $a=-0.8$, $H=0.5$, $h=0.7$ and
variable $\gamma $ to study if knowledge of singular points permits prediction of 
emergence of $1:2$ resonances. 

\begin{figure}[h!]
\center
\includegraphics[width=6cm, height=6cm]{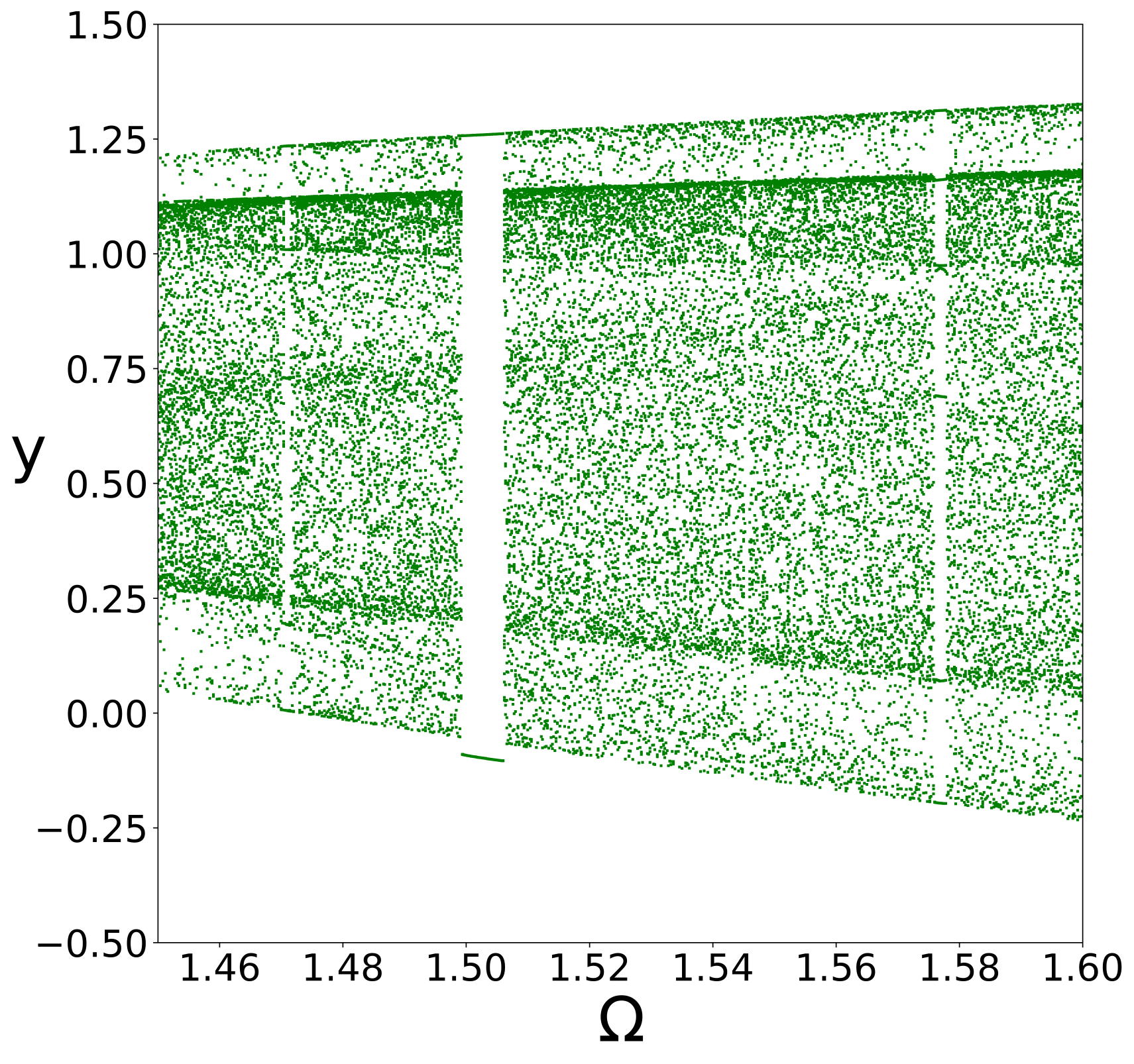}
\includegraphics[width=6cm, height=6cm]{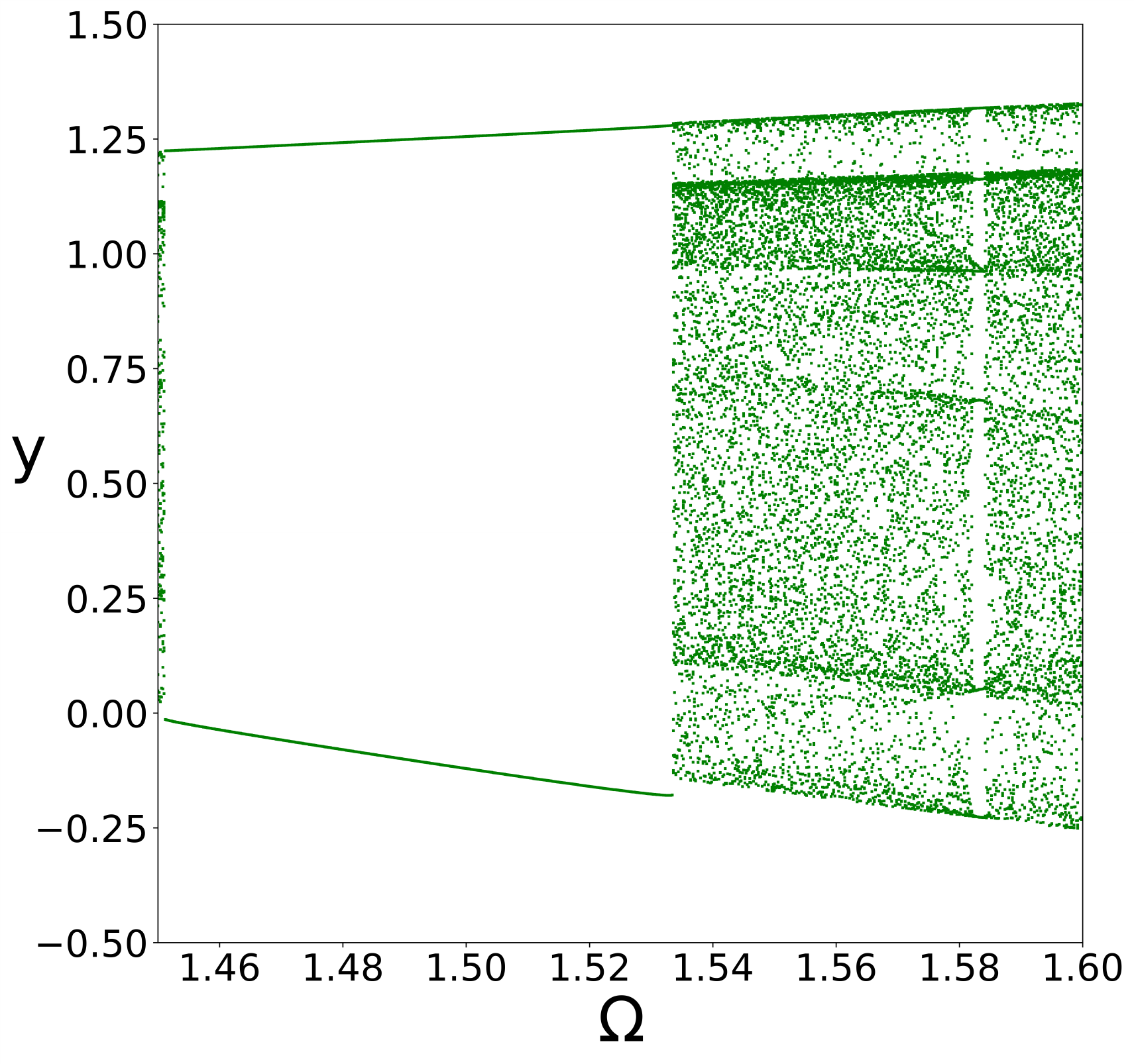}
\caption{Bifurcation diagrams: $\gamma = 2.773\,6$ -- left figure, $\gamma = 2.75$ -- right figure.
}
\label{F5}
\end{figure}

Left-hand figure \ref{F5} shows birth of $1:2$  resonance out from chaos. Right-hand figure  \ref{F5} 
displays a fully developed  $1:2$ resonance. The resonance appears at  $\gamma = 4.75$, in 
qualitative agreement with the computed value  $\gamma =5.\,375\,442$. 

\begin{figure}[h!]
\center
\includegraphics[width=11cm, height=7cm]{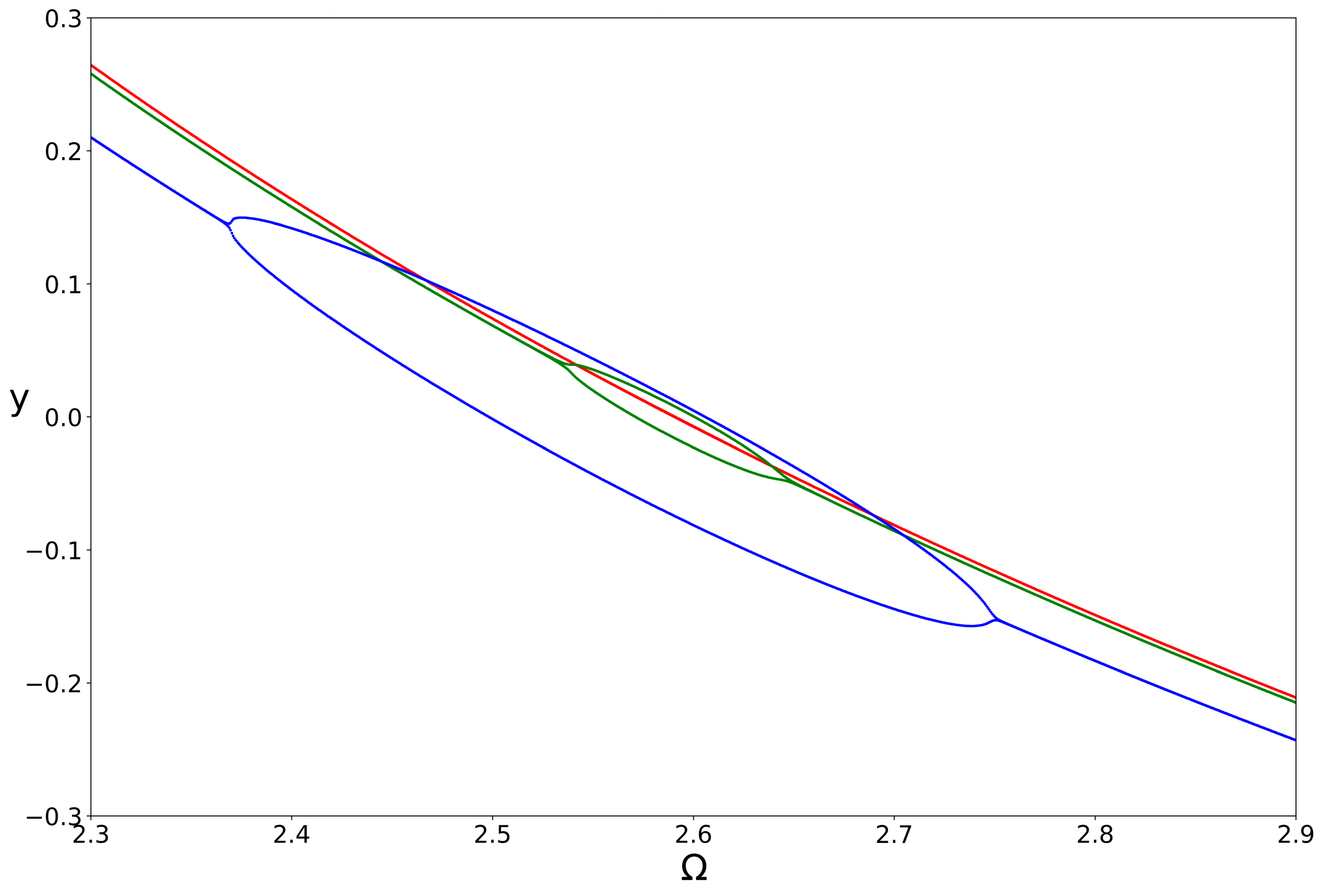}
\caption{Bifurcation diagram, $\gamma = 4.75$ -- Red, $\gamma = 4.77$ -- Green, $\gamma = 4.60$ -- Blue. 
}
\label{F6}
\end{figure}

Fig. \ref{F6} describes period doubling of $1:1$ resonance. Red curve corresponds to 
the $1:1$ resonance just before the first period doubling at $\gamma > 4.751$. 
Green and blue curves show growth of the $1:2$ resonance (before the next period doubling). 

\section{Conclusions}
We have demonstrated that on the basis of asymptotic solution(\ref{SOL}) to the effective equation (\ref{eff}) 
(Duffing equation is a special case) the birth of $1:2$ resonances can be predicted. 

 More precisely, implicit function \ref{F(Omega,A)}, computed 
from Eqs. (\ref{SOL}), has singular isolated points -- fulfilling Eqs. (\ref{SING1}) -- corresponding to the birth of 
$1:2$ resonances. Singular isolated points are computed as follows: 1. values of $h$, $a$, $H$ are chosen, 
2. equations  (\ref{SING1}) are solved numerically yielding many solutions -- values of $\Omega$, $A$, $\gamma$, 
3. real positive solutions are selected, 4. isolated points are found -- 
in such points $\frac{\partial ^{2}F}{\partial \Omega ^{2}}\frac{\partial
^{2}F}{\partial A^{2}}-\left( \frac{\partial ^{2}F}{\partial \Omega \partial
A}\right) ^{2}>0$. 

There are two kinds of such singular isolated points, (i) with $A\neq 0$, (ii) with $A=0$, 
which are solutions of simpler equations   (\ref{SING2}) which can be simplified further; see 
Eqs.  (\ref{p}), (\ref{q}). 
Singular isolated points of the first kind ($A\neq 0$) correspond to birth of $1:2$ resonance without 
contact with the primary $1:1$ resonance. Interestingly, $1:2$ resonance appears in the 
chaotic regime. 
On the other hand, singular points of the second kind ($A=0$) 
represent emergence of $1:2$ resonance due to period doubling of the primary resonance.

Singular isolated points computed from  Eqs. (\ref{SING1}) are very helpful in the search for the birth of $1:2$ resonances 
when solving  Eq. \ref{eff} numerically, although the  agreement is only qualitative (it can be improved 
 upon adding in Eq. (\ref{resonance}) the second harmonic). 

Since the effective equation (\ref{eff}) approximates well the system of coupled oscillators \cite{Okninski2006}, 
we expect that our findings apply also to the general model (\ref{general}).

%
%
%
%

\appendix{}
\section{Simplifying equations (\ref{SING2})}
\label{SE}

\setcounter{equation}{0}
\numberwithin{equation}{section}

Equations (\ref{SING2}), involving a complicated function $F\left( \Omega ,0;h,a,H,\gamma \right)$,  
can be simplified. More precisely, they can be reduced to polynomial equations (\ref{p}), (\ref{q})
\begin{equation}
p\left( x,\Omega ,a,H\right) =a_{4}\left( \Omega ,a,H\right)
x^{4}+a_{2}\left( \Omega ,a,H\right) x^{2}+a_{0}\left( \Omega ,a,H\right) =0
\label{p}
\end{equation}%
where $x=\gamma \frac{\Omega ^{2}}{1+\Omega ^{2}}$ and coefficients $a_{4}$, 
$a_{2}$, and $a_{0}$ are given below

\begin{tabular}{|l|}
\hline
$a_{4}\left( \Omega ,a,H\right) =\sum\nolimits_{k=0}^{3}c_{k}\Omega ^{2k}$
\\ \hline\hline
$c_{3}=1800$ \\ \hline
$c_{2}=-2160a+360+1080H^{2}$ \\ \hline
$c_{1}=-360H^{2}+720a+360a^{2}$ \\ \hline
$c_{0}=-360H^{2}+720a+360a^{2}$ \\ \hline
\end{tabular}

\medskip

\begin{tabular}{|l|}
\hline
$a_{2}\left( \Omega ,a,H\right) =\sum\nolimits_{k=0}^{6}c_{k}\Omega ^{2k}$
\\ \hline\hline
$c_{6}=48$ \\ \hline
$c_{5}=72H^{2}-1008-144a$ \\ \hline
$c_{4}=144a^{2}+3408a+24H^{4}-96aH^{2}-1704H^{2}-336$ \\ \hline
$c_{3}=2784aH^{2}-336H^{2}+672a+24a^{2}H^{2}-48a^{3}-696H^{4}-4176a^{2}$ \\ 
\hline
$c_{2}=-1080a^{2}H^{2}+2160a^{3}$ \\ \hline
$c_{1}=-672a^{3}-384a^{4}+336a^{2}H^{2}$ \\ \hline
$c_{0}=336a^{4}$ \\ \hline
\end{tabular}

\medskip

\begin{tabular}{|l|}
\hline
$a_{0}\left( \Omega ,a,H\right) =\sum\nolimits_{k=0}^{9}c_{k}\Omega ^{2k}$
\\ \hline\hline
$c_{9}=-1$ \\ \hline
$c_{8}=-1+6a-3H^{2}$ \\ \hline
$c_{7}=-15a^{2}+6a-3H^{2}-3H^{4}+64+12aH^{2}$ \\ \hline
$c_{6}=\left( 
\begin{array}{l}
-15a^{2}-18a^{2}H^{2}+6H^{4}a+192H^{2}+64 \\ 
+20a^{3}+12aH^{2}-384a-H^{6}-3H^{4}%
\end{array}%
\right) $ \\ \hline
$c_{5}=\left( 
\begin{array}{l}
-15a^{4}-384a+192H^{2}-18a^{2}H^{2}+20a^{3}+6H^{4}a \\ 
-768aH^{2}-H^{6}+192H^{4}+12a^{3}H^{2}-3a^{2}H^{4}+960a^{2}%
\end{array}%
\right) $ \\ \hline
$c_{4}=\left( 
\begin{array}{l}
-3a^{4}H^{2}-3a^{2}H^{4}+192H^{4}-768aH^{2}+64H^{6}+1152a^{2}H^{2} \\ 
+6a^{5}+12a^{3}H^{2}-15a^{4}-1280a^{3}+960a^{2}-384H^{4}a%
\end{array}%
\right) $ \\ \hline
$c_{3}=\left( 
\begin{array}{l}
64H^{6}+1152a^{2}H^{2}-384H^{4}a+6a^{5}-768a^{3}H^{2}+192a^{2}H^{4} \\ 
-1280a^{3}-3a^{4}H^{2}-a^{6}+960a^{4}%
\end{array}%
\right) $ \\ \hline
$c_{2}=-a^{6}-384a^{5}+960a^{4}+192a^{2}H^{4}-768a^{3}H^{2}+192a^{4}H^{2}$
\\ \hline
$c_{1}=-384a^{5}+192a^{4}H^{2}+64a^{6}$ \\ \hline
$c_{0}=64a^{6}\allowbreak $ \\ \hline
\end{tabular}

\bigskip 
\begin{equation}
q\left( y,x,\Omega ,a,H\right) =b_{2}\left( \Omega ,a,H\right)
y^{2}+b_{0}\left( x,\Omega ,a,H\right)  \label{q}
\end{equation}%
where $y=h\Omega $, $x=\gamma \frac{\Omega ^{2}}{1+\Omega ^{2}}$ is a
solution of Eq. (\ref{p}), and coefficients $b_{2}$, $b_{0}$ are provided
below

\begin{tabular}{|l|}
\hline
$b_{2}\left( \Omega ,a,H\right) =\sum\nolimits_{k=0}^{5}c_{k}\Omega ^{2k}$
\\ \hline\hline
$c_{5}=10$ \\ \hline
$c_{4}=-32a+16H^{2}+2$ \\ \hline
$c_{3}=6H^{4}-24aH^{2}+36a^{2}$ \\ \hline
$c_{2}=-16a^{3}+8aH^{2}+8a^{2}H^{2}-12a^{2}-2H^{4}$ \\ \hline
$c_{1}=2a^{4}+16a^{3}-8a^{2}H^{2}$ \\ \hline
$c_{0}=-6a^{4}$ \\ \hline
\end{tabular}

$\medskip $

\begin{tabular}{|l|}
\hline
$b_{0}\left( x,\Omega ,a,H\right) =\sum\nolimits_{k=0}^{7}c_{k}\Omega ^{2k}$
\\ \hline\hline
$c_{7}=3$ \\ \hline
$c_{6}=-10a-39+5H^{2}$ \\ \hline
$c_{5}=126a+120-63H^{2}+12a^{2}+2H^{4}-8aH^{2}$ \\ \hline
$c_{4}=192H^{2}-24H^{4}+36x^{2}-6a^{3}+96aH^{2}-144a^{2}+3a^{2}H^{2}-384a$
\\ \hline
$c_{3}=\left( 
\begin{array}{l}
-64H^{2}+72H^{4}+432a^{2}+a^{4}-33a^{2}H^{2}-48ax^{2}+128a \\ 
-324x^{2}-288aH^{2}+24H^{2}x^{2}+66a^{3}%
\end{array}%
\right) $ \\ \hline
$c_{2}=\left( 
\begin{array}{l}
336ax^{2}-192a^{3}+96a^{2}H^{2}-168H^{2}x^{2}-64H^{4}-384a^{2} \\ 
-9a^{4}+12x^{2}a^{2}+256aH^{2}%
\end{array}%
\right) $ \\ \hline
$c_{1}=-192a^{2}H^{2}+168H^{2}x^{2}-336ax^{2}-12x^{2}a^{2}+24a^{4}+384a^{3}$
\\ \hline
$c_{0}=-128a^{4}+336x^{2}a^{2}$ \\ \hline
\end{tabular}

\section{Computational details}

Nonlinear polynomial equations were solved numerically using Maple's computational engine from Scientific WorkPlace 4.0. All Figures were
plotted with the computational engine MuPAD from Scientific WorkPlace 5.5.
Curves shown in bifurcation diagrams in Figs. \ref{F2}, \ref{F3}, \ref{F5}, \ref{F6} 
were computed running DYNAMICS, a program written by Helena E. Nusse and James 
A. Yorke \cite{Nusse2012}, and our programs written in Pascal and Python \cite{Perez2007}.


\begin{thebibliography}{99}
\bibitem{Kyziol2024}  J. Kyzio{\l }, A. Okni\'{n}ski, Asymmetric Duffing oscillator: 
metamorphoses of $1:2$ resonance and its interaction with the primary resonance, eprint \textit{%
arXiv:}2407.03423v2 [nlin.CD].

\bibitem{Mahmoud2004} G.M. Mahmoud, T. Bountis, The dynamics of systems of
complex nonlinear oscillators: a review, \textit{Int. J. Bifur. Chaos} 
\textbf{14} (2004) 3821-3846.

\bibitem{Pikovsky2015} A. Pikovsky, M. Rosenblum, Dynamics of globally
coupled oscillators: Progress and perspectives, \textit{Chaos}, \textbf{25}
(2015) 097616.

\bibitem{Schultheiss2011} N.W. Schultheiss, A.A. Prinz, R.J. Butera, eds. 
\textit{Phase response curves in neuroscience: theory, experiment, and
analysis}. Springer A Science \& Business Media, 2011.

\bibitem{Awal2019} N.M. Awal, D. Bullara, I.R. Epstein, The smallest
chimera, Periodicity and chaos in a pair of coupled chemical oscillators, 
\textit{Chaos} \textbf{29} (2019) 013131.

\bibitem{Hajjaj2019} A.Z. Hajjaj, N. Jaber, S. Ilyas, F.K. Alfosail, M.I.
Younis, Linear and nonlinear dynamics of micro- and nano-resonators: Review
of recent advances, \textit{Int. J. Non-Linear Mechanics}, 2019, March 2020,
103328, https://doi.org/10.1016/j.ijnonlinmec.2019.103328.

\bibitem{Kozlowski1995} J. Koz{\l }owski, U. Parlitz and W. Lauterborn,
Bifurcation analysis of two coupled periodically driven Duffing oscillators, 
\textit{Phys. Rev.} E \textbf{51} (1995) 1861-1867.

\bibitem{Kuznetsov2009} A. P. Kuznetsov, N. V. Stankevich and L. V Turukina,
Coupled van der Pol-Duffing oscillators: phase dynamics and structure of
synchronization tongues, \textit{Physica} D \textbf{238} (2009) 1203-1215.

\bibitem{Wiercigroch2009} N.N. Verichev, S.N. Verichev, M. Wiercigroch,
C-oscillators and stability of stationary cluster structures in lattices of
diffusively coupled oscillators, \textit{Chaos, Solitons} \& \textit{Fractals%
} \textbf{42} (2009) 686-701.

\bibitem{Perkins2012} E. Perkins, B. Balachandran, Noise-enhanced response
of nonlinear oscillators, \textit{Procedia Iutam} \textbf{5} (2012) 59-68.

\bibitem{Sabrathinam2013} S. Sabarathinam, K. Thamilmaran, L. Borkows ki, P.
Perlikowski, P. Brzeski, A. Stefanski, T. Kapitaniak, Transient chaos in two
coupled, dissipatively perturbed Hamiltonian Duffing oscillators, \textit{%
Commun. Nonlinear Sci Numer. Simulat.} \textbf{18} (2013) 3098-3107.

\bibitem{Zulli2016} D. Zulli, A. Luongo, Control of primary and subharmonic
resonances of a Duffing oscillator via non-linear energy sink, \textit{Int.
J. Non-Linear Mechanics} \textbf{80} (2016) 170-182.

\bibitem{Luo2017} B. Yu, A.C.J. Luo, Analytical period-1 motions to chaos in
a two-degree-of-freedom oscillator with a hardening nonlinear spring, 
\textit{Int. J. Dynam. Control} \textbf{5} (2017) 436--453.

\bibitem{Karahan2017} M.M. Faith Karahan, M. Pakdemirli, Free and forced
vibrations of the strongly nonlinear cubic-quintic Duffing oscillators, 
\textit{Zeit. Natur.} A \textbf{72} (2017) 59-69.

\bibitem{Papangelo2019} A. Papangelo, F. Fontanela, A. Grolet, M.
Ciavarella, N. Hoffmann, Multistability and localization in forced cyclis
structures modelled by weakly-coupled Duffing oscillators, \textit{J. Sound
Vibr.} \textbf{440} (2019) 202-211.

\bibitem{DenHartog1985} J. P. Den Hartog, \textit{Mechanical Vibrations}
(4th edition), Dover Publications, New York 1985.

\bibitem{Oueini1999} S. S. Oueini, A. H. Nayfeh and J.R. Pratt, A review of
development and implementation of an active nonlinear vibration absorber%
\textit{Arch. Appl. Mech.}, \textbf{69} (1999) 585-620.

\bibitem{Okninski2006} A.~Okni\'{n}ski and J.~Kyzio\l, Perturbation
analysis of the effective equation for two coupled periodically driven
oscillators, \textit{Diff. Eqs. Nonlin. Mech.} \textbf{2006} (2006) 56146.

\bibitem{Kyziol2013b} J.~Kyzio\l\ and A.~Okni\'{n}ski, Exact nonlinear
fourth-order equation for two coupled oscillators: metamorphoses of
resonance curves, \textit{Acta Phys. Polon.} B \textbf{44} (2013) 35-47.

\bibitem{Nayfeh2011} A.H.~Nayfeh, \textit{Introduction to Perturbation
Techniques}, John Wiley \& Sons, 2011.

\bibitem{Janickia} K.L. Janicki, W. Szempli\'{n}ska-Stupnicka. SUBHARMONIC
RESONANCES AND CRITERIA FOR ESCAPE AND CHAOS IN A DRIVEN OSCILLATOR, \textit{%
Journal of Sound and Vibration} \textbf{180} (1995) 253-269.

\bibitem{Janickib} K.L. Janicki. PhD Thesis (in Polish), Institute of
Fundamental Technological Research PAS, 16/1994, 
\url{
https://rcin.org.pl/dlibra/doccontent?id=686}

\bibitem{Fikhtengolts2014} G.M. Fikhtengol'ts, (I.N Sneddon, Editor) \textit{%
The fundamentals of mathematical analysis}, Vol. 2, Elsevier, 2014 (Chapter
19), translated from Russian, Moscow, 1969.

\bibitem{Kyziol2024a} J. Kyzio\l , A. Okni\'{n}ski. The Twin-Well Duffing
Equation: Escape Phenomena, Bistability, Jumps, and Other Bifurcations, 
\textit{Nonlinear Dynamics and Systems Theory} \textbf{24} (2024) 181-192.

\bibitem{Nusse2012} Nusse, Helena E., James A. Yorke. \textit{Dynamics:
numerical explorations: accompanying computer program dynamics.} Vol. 101.
Springer, 2012.

\bibitem{Perez2007} Fernando P\'{e}rez, Brian E. Granger, IPython, \textit{A
System for Interactive Scientific Computing}, Computing in Science and
Engineering, vol. 9, no. 3, pp. 21-29, May/June 2007,
doi:10.1109/MCSE.2007.53. 
\url{
https://ipython.org}






%
%
%
%
%
%



\end{thebibliography}
\end{document}